\documentclass[prb,aps,showpacs,prb,amsmath,floatfix,showkeys]{revtex4}
\usepackage[]{graphicx}
\usepackage{graphicx}
\usepackage{dcolumn}
\usepackage{bm}

\begin{document}
\title{Kondo resonance in an ac driven quantum dot subjected
to finite bias}
\author{A. Goker$^{1,2}$}

\affiliation{$^1$
Department de Chimie, Universite de Montreal, \\
C.P.6128 Succursale A, Montreal, Quebec H3C 3J7, Canada
}

\affiliation{$^2$
Department of Physics, Fatih University, \\
Buyukcekmece, Istanbul 34500, Turkey 
}

\date{\today}


\begin{abstract}
We employ the time-dependent non-crossing approximation to study 
the time averaged conductance for a single electron transistor in
the Kondo regime when the dot level is sinusoidally
driven from its equilibrium position by means
of a gate voltage in finite bias. We find that the average
conductance exhibits significant deviation from the monotonous reduction when
the applied bias is equal to the driving frequency of the dot level.
We investigate the effect of the temperature and the driving frequency
on the observed enhancement. We attribute this behaviour to
the overlap of the satellite Kondo peaks with the split Kondo
resonances formed at each lead's Fermi level. We display the
spectral function to put our interpretation into more rigorous footing.  
\end{abstract}



\maketitle

The advent of the state-of-the-art nanotechnology experiments 
provided unprecedented control over nanostructures like
quantum dots which enabled to test various theoretical
predictions. One of these predictions was a long standing
question on whether the Kondo effect, which is a hallmark
of condensed matter physics, would dominate the low temperature
transport through a quantum dot \cite{NgLee88PRL,GlazmanRaikh88JL}. 
Subsequent experiments resolved this controversy by confirming these 
predictions \cite{GoldhaberetAl98Nature,GoldhaberetAl98PRL,
CronenwettetAl98Science}.    
 
The Kondo effect derives its name from the seminal work 
of Jun Kondo \cite{Kondo64PTP} in which he discovered that bulk 
metals with magnetic impurities providing localized unpaired spins
would experience an enhancement in their resistivities at low
temperatures due to the formation of a spin singlet resulting 
from the interaction of the unpaired localized electron and 
delocalized electrons near the Fermi level \cite{KouwenhovenGlazman01PW}. 
The requirement to obtain the Kondo effect in quantum dots is 
to confine odd number of electrons in the dot. This results in 
a net spin in the dot and a new transport channel opens when this 
net spin is coupled to the fermionic bath in the metallic leads 
giving rise to a peak at the Fermi level in the dot density of states. 
Ambient temperature, bias and magnetic field greatly influence the 
sharpness of this peak

The effects of abrupt perturbations by step-like switching 
of the gate or source-drain voltage have been investigated 
extensively \cite{NordlanderetAl99PRL,PlihaletAl00PRB,
SchillerHershfield00PRB,MerinoMarston04PRB} and they 
unambiguously demonstrated that the resulting transient 
current exhibits three different time scales 
\cite{PlihaletAl05PRB,AndersSchiller05PRL,AndersetAl06PRB,
IzmaylovetAl06JPCM}. The initial fast non-Kondo timescale
is characterized by the reshaping of the Fano resonance 
while the Kondo resonance reaches a metastable state at the
end of the much longer Kondo timescale. A third and longest 
timescale emerges when the Kondo resonance splits for finite
source drain bias. Later studies predicted that the asymmetric
coupling of the dot to the contacts may induce interference
between the Kondo resonance and the discontinuities in the
density of states of the leads \cite{GokeretAl07JPCM}. 

Time-dependent AC perturbations by step-like switching of the
gate or source-drain voltages provide another intriguing
way to probe the single-electron transport in the Kondo
regime since the nonadiabatic photon-assisted
processes interfere with the Kondo tunneling
giving rise to sidebands of the main Kondo resonance
\cite{HettlerSchoeller95PRL,SchillerHershfield96PRL,
GoldinAvishai98PRL,KaminskyetAl99PRL,NordlanderetAl00PRB,
KoganetAl04Science}.
More recently, precise control over the dot-lead
coupling \cite{ScheibleetAl04PRL,ParksetAl07PRL} enabled 
the study of quite exotic phenomenon like Kondo shuttling in 
which the Kondo temperature fluctuates periodically due to 
the oscillation of the central island between the leads
\cite{AlhassaniehetAl05PRL,KiselevetAl06PRB}.

In this paper, we will consider a slightly 
different scenario in which an applied ac gate voltage
drives the dot level sinusoidally from its equilibrium
level in the Kondo regime while a finite bias is applied 
to the source and drain contacts. This causes periodical
oscillations of the Kondo temperature and thus the instantaneous
conductance. Splitting of the Kondo peak and the formation of
the Kondo sidebands take place simultaneously. We report
the resulting time averaged conductance arising from the
presence of these two phenomena.

We model this device by a single spin degenerate level
of energy $\epsilon_{dot}$ attached to leads through tunnel
barriers. Single impurity Anderson Hamiltonian governs the 
physics of this system quite adequately. We carry out the 
auxiliary boson transformation for the Anderson Hamiltonian 
where the ordinary electron operator on the impurity is rewritten
in terms of a massless boson operator and a pseudofermion
operator. $U$ $\rightarrow \infty$ limit is obtained by imposing the 
condition that the sum of the number of bosons and the
pseudofermions is equal to unity. The resulting Hamiltonian
is given by
\begin{equation}
H(t)=\sum_{\sigma}\epsilon_{dot}(t)n_{\sigma}
+\sum_{k\alpha\sigma}\left [\epsilon_{k}n_{k\alpha\sigma}+
V_{\alpha}(\varepsilon_{k\alpha},t)c_{k\alpha\sigma}^{\dag}
b^{\dag}f_{\sigma}+{\rm H.c.} \right],
\end{equation}
where $f_{\sigma}^{\dag}(f_{\sigma})$ and $c_{k\alpha\sigma}^{dag}(c_{k\alpha\sigma})$
with $\alpha$=L,R create(annihilate) an electron of spin $\sigma$ in the impurity
and in the left(L) and right(R) contacts respectively. The $n_{\sigma}$ and
$n_{k\alpha\sigma}$ are the corresponding number operators and $V_{\alpha}$ are the
hopping amplitudes for the left and the right leads. $b^{\dag}(b)$ creates(annihilates)
a slave boson in the impurity. In this paper, we will adopt atomic units 
with $\hbar=k_B=e=1$.

We will also assume that the hopping matrix elements are equal with
no explicit time and energy dependency while the density of states
in both contacts are parabolic with the same bandwidth. In this
case, the coupling of the quantum dot to the leads can be written as
$\Gamma(\epsilon)=\bar{\Gamma}\rho(\epsilon)$ where $\bar{\Gamma}$ 
is a constant given by $\bar{\Gamma}=2\pi|V(\epsilon_f)|^2$ and
$\rho(\epsilon)$ is the DOS function.

We invoke the well tested non-crossing approximation(NCA) to obtain
the pseudofermion and slave boson self-energies and then solve
the resulting real-time coupled integro-differential Dyson equations
for the retarded and less than Green's functions in a discrete
two-dimensional grid. A technical description of our implementation
have been published elsewhere \cite{ShaoetAl194PRB,IzmaylovetAl06JPCM}.

The net current flowing in the circuit can readily be calculated          
from the resulting Green's functions $G_{pseu}^{<(R)}(t,t')$ and
$B^{<(R)}(t,t')$. We will denote the net current with
$I(t)=I_L(t)-I_R(t)$, where $I_L(t)(I_R(t))$ represents the 
net current from the left(right) contact through the 
left(right) barrier to the quantum dot. The general expression
for the net current \cite{JauhoetAl94PRB} can be rewritten by using 
pseudofermion and slave boson Green's functions \cite{GokeretAl07JPCM}. 
The final expression for the current is
\begin{eqnarray}
I(t)&=&2\bar{\Gamma} Im(i \int_{-\infty}^{t} dt_1
(G_{pseu}^{R}(t,t_1)B^{<}(t_1,t)+G_{pseu}^{<}(t,t_1)B^{R}(t_1,t)) \nonumber \\
& &(f_{L}(t-t_1)-f_{R}(t-t_1))),
\end{eqnarray}
where $f_L(t-t_1)$ and $f_R(t-t_1)$ are the convolution of the
density of states function with the Fermi-Dirac distribution \cite{GokeretAl07JPCM}.
The conductance $G$ is given by the current divided by the bias voltage.

A quantum coherent many-body state called the Kondo effect emerges
when the dot level is situated below the Fermi energy at sufficiently
low temperatures. A spin singlet is formed from the free spin
localized in the dot and the Fermi sea of electrons in the contacts.
Its manifestation is a sharp resonance pinned to the Fermi levels 
of the contacts in the dot density of states. The linewidth of the 
Kondo resonance is well described by an energy scale $T_K$ 
(Kondo temperature) given by
\begin{equation}
T_K \propto \left(\frac{D\Gamma}{4}\right)^\frac{1}{2}
\exp\left(-\frac{\pi|\epsilon_{\rm dot}|}{\Gamma}\right),
\label{tkondo}
\end{equation}
where $D$ is a high energy cutoff equal to half bandwidth
of the conduction electrons and $\Gamma$ corresponds to
the value of coupling between the dot and the contacts 
$\Gamma(\epsilon)$ at $\epsilon=\epsilon_F$.

The purpose of this paper is to theoretically investigate 
a case in which the dot level is displaced from its equilibrium 
level sinusoidally by means of a gate voltage. This results in a 
de facto time dependent Kondo temperature and causes the
instantaneous conductance to exhibit periodic modulations
with a minimum when the dot level is farthest from the
Fermi level and a maximum when the dot level is closest. 
The period of conductance oscillations is equal to the driving 
frequency. In particular, we will investigate a system which has 
been analyzed in zero bias before. Its dot level is given by 
$\epsilon_{dot}(t)$=-5$\Gamma$+4$\Gamma$ cos $\Omega t$. We are
going to report the conductance results time averaged over a full
period of oscillation. 

\begin{figure}[htb]
\centerline{\includegraphics[angle=90,width=7.5cm,height=6cm]{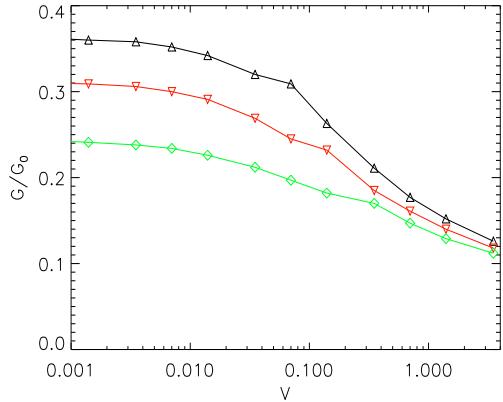}}
\caption{
Black(upward triangles), red(downward triangles) and green(diamonds)
curves show time-averaged conductance results as a function of 
source-drain voltage for $\Omega$=0.07$\Gamma$, $\Omega$=0.14$\Gamma$
and $\Omega$=0.35$\Gamma$ respectively at T=0.005$\Gamma$.
}
\label{Fig1}
\end{figure}

Fig.~\ref{Fig1} depicts the time averaged conductance as a function
of the source-drain bias for three different driving frequencies.
In each case, the calculation starts from the zero bias limit
reported previously \cite{NordlanderetAl00PRB}. Driving with a lower 
frequency results in a larger conductance since the Kondo resonance 
can find more time to form during the time the dot level lies above 
its equilibrium position. As the bias is gradually increased, the 
conductance starts to decrease since the Kondo resonance 
starts to be quenched. However, Fig.~\ref{Fig1} clearly shows that
the conductance is enhanced considerably when the driving frequency
is equal to the applied source-drain bias, deviating from
the monotonous decreasing behaviour.

\begin{figure}[htb]
\centerline{\includegraphics[angle=90,width=7.5cm,height=6cm]{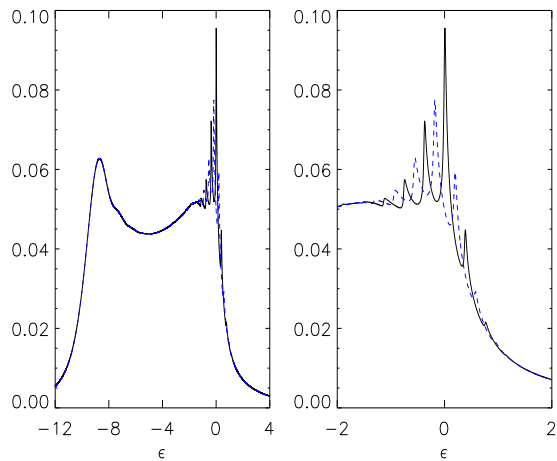}}
\caption{
Black(full) curve displays the time averaged dot density 
of states for $\Omega$=0.35$\Gamma$ at T=0.005$\Gamma$ in
infinitesimal bias while blue(dashed) curve shows the dot
density of states for V=$\Omega$=0.35$\Gamma$ at T=0.005$\Gamma$
in the left panel. Right panel is the magnification of the left
panel around the Fermi level. Both the split Kondo peaks
and the satellite Kondo resonances are clearly visible here. 
}
\label{spectral}
\end{figure}

The origin of this phenomenon can be explained by noting that
the finite bias induces a splitting of the Kondo resonance
pinned to the Fermi level into two distinct Kondo peaks
each located at the left and right reservoir's Fermi levels
and these split Kondo peaks are separated by the applied
source-drain voltage. Furthermore, previous studies 
demonstrated that the ac driving generates sidebands of
the split Kondo peaks due to the interference between 
the nonadiabatic photon-assisted processes and the Kondo
tunneling. The separation of the satellite Kondo
resonances from each split Kondo peak is an integer
times the driving frequency $\Omega$. Therefore, when 
the driving frequency is equal to the source-drain bias,
the first satellite resonance of each split Kondo peak 
overlaps with the other split Kondo peak. We believe 
this is the reason of the enhancement in the conductance.
Fig.~\ref{spectral} displays all these features in the
time averaged dot density of states, which is obtained 
by averaging the instantaneous dot density of states
in a full period of oscillation. The instantaneous dot
density of states is the Fourier transform of the retarded
Green's function and it is given by
\begin{equation}
A(\epsilon,t)=\frac{1}{2\pi}\int_{-\infty}^{t} dt_1
(G_{pseu}^{R}(t,t_1)B^{<}(t_1,t)+G_{pseu}^{<}(t,t_1)B^{R}(t_1,t))
e^{i \epsilon t_1}.
\end{equation} 
The two broad peaks at $\epsilon$=-1$\Gamma$ and $\epsilon$=-9$\Gamma$
in the left panel correspond to the stationary states of the oscillation
whereas the dip at $\epsilon$=-5$\Gamma$ is the equilibrium position
of the dot. Right panel shows magnification of the left one around
the Fermi level both at infinitesimal bias and when V=$\Omega$. All the
features described above are visible here quite clearly.

\begin{figure}[htb]
\centerline{\includegraphics[angle=90,width=7.5cm,height=6cm]{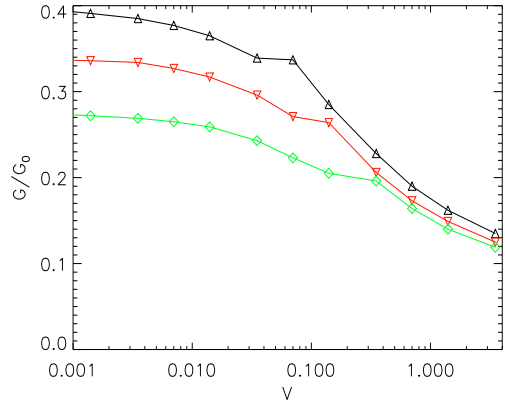}}  
\caption{
Black(upward triangles), red(downward triangles) and green(diamonds)
curves show time-averaged conductance results as a function of
source-drain voltage for $\Omega$=0.07$\Gamma$, $\Omega$=0.14$\Gamma$
and $\Omega$=0.35$\Gamma$ respectively at T=0.002$\Gamma$.
}
\label{Fig3}
\end{figure}

If the above hypothesis really holds, the enhancement would
strongly depend on the ambient temperature. In order to verify
the veracity of this argument, we calculated the conductance
at the same frequencies in Fig.~\ref{Fig1} for a lower 
temperature. The results are shown in Fig.~\ref{Fig3}. The
enhancements at each frequency indeed increase compared
to the higher temperature. Split Kondo peaks are more developed
at lower temperature and so are their satellite resonances thus
their overlap is more substantial giving rise to larger
enhancement.

\begin{figure}[htb]
\centerline{\includegraphics[angle=90,width=7.5cm,height=6cm]{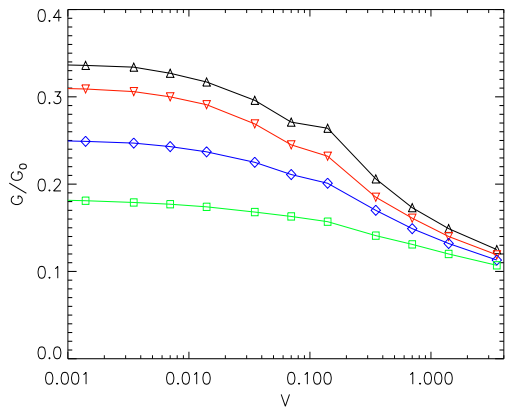}}
\caption{
Black(upward triangles), red(downward triangles), blue(diamonds)
and green(squares) curves show time-averaged conductance results as 
a function of source-drain voltage for T=0.002$\Gamma$, 
T=0.005$\Gamma$, T=0.02$\Gamma$ and T=0.1$\Gamma$ respectively 
at $\Omega$=0.14$\Gamma$.
}
\label{Fig4}
\end{figure}

Finally, Fig.~\ref{Fig4} shows the behaviour of the observed
enhancement for a fixed oscillation frequency and various 
temperatures as a function of the source-drain bias. The
enhancement is the largest at the lowest temperature since
both the split and satellite Kondo peaks are fully
developed. When we start decreasing the temperature the 
conductance at zero bias gets suppressed alongside with
the enhancement at V=$\Omega$ as the Kondo effect is
gradually quenched. At the highest temperature the
enhancement becomes indiscernible. The Kondo resonance
even at zero bias is not developed sufficiently at this
temperature hence the reduction of the conductance as a 
function of the bias is not as dramatic as lower temperatures.
Therefore, the overlap of the sidebands with the split peaks
creates negligible enhancement.  

In conclusion, the non-crossing approximation was employed 
to investigate the effect of finite bias on time averaged
conductance in single electron transistor when the dot level is
driven sinusoidally by means of a gate voltage. Our results 
indicate that the time averaged conductance displays significant
enhancement from its monotonous decrease when the driving frequency
is equal to the applied source-drain bias. The enhancement tends to 
increase with decreasing temperature. We attributed this
behaviour to the overlap of the split Kondo peaks pinned to each
contact's Fermi level with the satellite Kondo resonances. We
expect this effect to be observable with present day experimental
techniques which can directly probe current voltage characteristics.
We thus hope to motivate new experiments in this field with this
Brief Report. 

A.G is extremely grateful to NSERC for providing generous financial 
support.


\bibliographystyle{iopams}
\bibliography{gen}

\end{document}